\documentclass[twocolumn,prd,aps,nofootinbib,showpacs,superscriptaddress]{revtex4}
\usepackage{multirow}
\usepackage{graphicx}
\usepackage[dvips]{color}
\usepackage{amssymb,amsmath}

\begin{document}
%////////////////////////////////////////////////////////////////////

%====================================================================
\title{On the fraction-dimension migration of self-interstitials in zirconium}
%====================================================================

%--------------------------------------------------------------------
\author{Rui Zhong}
\author{Chaoqiong Ma}
\author{Baoqin Fu}
\author{Jun Wang}
\author{Qing Hou}
%--------------------------------------------------------------------
\email{qhou@scu.edu.cn}
\affiliation{
Key Laboratory for Radiation Physics and Technology of  Education Ministry;
Institute of Nuclear Science and Technology,
Sichuan University, Chengdu 610064, P. R. China
}
%================================================================
\date{\today}
%================================================================

%----------------------------------------------------------------
\begin{abstract}
Molecular dynamic simulations were conducted to study the self-interstitial migration in zirconium. By defining the crystal lattice point, at which more than one atom fall in the Wigner-Seitz cell of the lattice point, for the location of interstitial atoms (LSIA), three types of events were identified for LSIA migration: the jump remaining in one $\langle11\overline{2}0\rangle$ direction (ILJ), the jump from one $\langle11\overline{2}0\rangle$ to another $\langle11\overline{2}0\rangle$ direction in the same basal plane (OLJ) and the jump from one basal  plane to an adjacent basal plane (OPJ). The occurrence frequencies of the three types were calculated. ILJ was found to be the dominant event in the temperature range ($300 K $ to $1200 K$), but the occurrence frequencies of OLJ and OPJ increased with increasing temperature. Although the three types of jumps may not follow Brownian and Arrhenius behavior, on the whole, the migration of the LSIAs tend to be Brownian-like. Moreover, the migration trajectories of LSAs in the hcp basal-plane are not what are observed if only conventional one- or two-dimensional migrations exist; rather, they exhibit the feature we call fraction-dimensional. Namely, the trajectories are composed of line segments in $\langle11\overline{2}0\rangle$ directions with the average length of the line segments varying with the temperature. Using Monte Carlo simulations, the potential kinetic impacts of the fraction-dimensional migration, which is measured by the average number of lattice sites visited per jump event (denoted by $n_{SPE}$ ), was analyzed. The significant differences between the $n_{SPE}$ value of the fraction-dimensional migration and those of the conventional one- and two-dimensional migrations suggest that the conventional diffusion coefficient, which cannot reflect the feature of fraction-dimensional migration, cannot give an accurate description of the underlying kinetics of SIAs in Zr. This conclusion may not be limited to the SIA migration in Zr and could be more generally meaningful for situations in which the low dimensional migration of defects has been observed.
\end{abstract}
%----------------------------------------------------------------------


%---------------------------------------------
\maketitle

%////////////////////////////////////////////////////////////////////////////
%============================================================================
\section{Introduction}
%----------------------------------------------------------------------------
Because zirconium (Zr) has a small neutron absorption cross section, its alloys have been commonly used as fuel cladding and structural material in nuclear reactors. An intensively concerning issue herein is the degradation of material properties induced by prolonged exposure to the irradiation environment. The changes of material properties observed macroscopically originate from various atomistic processes \cite{Was:2007}. To bridge the macro observations with the micro processes, studies spanning multiple time and space scales are crucial \cite{Samaras:2009}. In the present paper, we focus on the migration of self-interstitial atoms (SIAs) in Zr. It is believed that the anisotropic migration of SIAs and vacancies, which can be produced by irradiation of energetic neutrons and ions as well as electrons, play essential roles in explaining the irradiation growth in Zr \cite{Wen:2012, Arevalo:2007, Barashev:2015, Woo:2007, Woo:1988, Semenov:2006}.

Simulation studies relevant to the migration behavior of SIAs in Zr have been reported by a number of groups using ab initio modeling or molecular dynamics \cite{Pasianot:2000, Osetsky:2002, Woo:2003, Domain:2006, Diego:2008, Diego:2011, Wen:2012, Verite:2013, Samolyuk:2014, Woo:2007, Varvenne:2013, Christensen:2015}. The most recent ab initio modeling showed that, in contrast to the symmetric octahedral configuration predicted earlier \cite{Domain:2006, Willaime:2003}, the lowest-energy SIA configuration should be the low-symmetric basal octahedral (BO) configuration \cite{Peng:2012, Samolyuk:2014, Verite:2013, Varvenne:2013}. In addition, it was found in \cite{Samolyuk:2014, Verite:2013, Varvenne:2013} that there may exist some low-symmetric metastable SIA configurations that were not recognized previously \cite{Willaime:1991, Peng:2012, Peng:2012}. These new results of ab initio modeling reveal that the migration of SIA in Zr should be anisotropic and involve complex atomistic processes. On the other hand, from the viewpoint of multiscale simulations, the extraction of the effective SIA diffusion coefficients of SIAs that condense the effects of the atomistic processes and more directly correlate to simulation and analysis methods of larger time-space scales is desired \cite{Christien:2005, Arevalo:2007, Barashev:2015, Samolyuk:2014, Peng:2012}. Using the nudged elastic band (NEB) method to obtain migration barriers between SIA configurations identified in ab initio modeling, Samolyuk et al estimated the SIA diffusion coefficients in the c-axis and in the basal plane of the hcp structure through event driving kinetics Monte Carlo (KMC) simulations \cite{Samolyuk:2014}. As noted by the authors, the out-of-plane diffusion coefficient was possibly overestimated because the same attempt jump frequency was assigned to all considered jump events. The jump frequencies can be calculated according to the formalism of the transition state theory of Vineyard \cite{Vineyard:1957}, in which the potential surface is assumed  to be harmonic in the neighboring region of local minimum and saddle points. At high temperatures, the anharmonicity of the potential surface may influence the calculated jump frequencies \cite{Wen:2012}. Thus, comparatively, molecular dynamic (MD) simulations provide a more intuitive approach to calculate the SIA diffusion coefficients or jump frequencies.

Using MD simulations, Pasianot et al calculated the SIA diffusion coefficients in the basal plane \cite{Pasianot:2000}. Based on the highly non-Arrhenius behavior of the obtained diffusion coefficients, the possibility that the SIAs perform one-dimensional movement in $\langle11\overline{2}0\rangle$ at low temperature was suggested. More detailed MD simulations conducted by Osetsky et al \cite{Osetsky:2002} provided further evidence for that the diffusion of SIAs was one-dimensional in the $\langle11\overline{2}0\rangle$ direction at low temperature and to change from one-dimensional to two-dimensional and then three-dimensional diffusion with increasing temperature. Shortly after Osetsky et al , Woo et al \cite{Woo:2003} published their MD results for SIA diffusion coefficient and also demonstrated the low-dimension diffusion of SIA. However, there was a remarkable difference in the basal diffusion coefficients given in these two works. The basal diffusion coefficients obtained in the former work exhibited good Arrhenius dependence on temperature at low temperatures, whereas the basal diffusion coefficients given in the later work depend on temperature very weakly. Given the same potential \cite{Ackland:1995} used in both works, the reason for this difference is unclear. More recent MD studies on the SIA diffusion in Zr were conducted by Mendelev and Bokstein \cite{Mendelev:2010} and Diego et al \cite{Diego:2011}. In these two studies, no obvious anisotropic diffusion was observed because of the used potential \cite{Mendelev:2007} leading to a symmetric most stable SIA configuration (octahedral configuration), a result that disagreed with the predictions of the newest ab initio modelling \cite{Peng:2012, Samolyuk:2014, Verite:2013, Varvenne:2013}.

In addition to the studies mentioned above, more detailed pictures of the SIA migration are needed. In the present paper, based on MD and MC simulations, we will demonstrate that the migration of SIAs in Zr tends to be Brownian-like and fraction-dimensional; namely, the migration trajectories of SIAs are constituted with a sequence of line segments lying in $\{0001\}$ planes, with the average length of the line segments being strongly temperature dependent. We will also demonstrate that this migration feature may have significant kinetic impacts, which cannot be reflected by the diffusion coefficient that is usually used to characterize the diffusion phenomenon.

%%%%%%%%%%%%%%%%%%%%%%%%%%%%%%%%%%%%%%%%%%%%%%%%%%%%%%%%%%%%%%%%%%%%%%%%%%%%%%%%
%===============================================================================
\section{MD Simulation of SIA migration}
\label{sec:MD simulation}
%-------------------------------------------------------------------------------
\subsection{Methods}
\label{subsec:Method}
Our MD simulations were performed using the graphics processing unit (GPU)-based MD package MDPSCU \cite{Hou:2013}. We used the many-body semi-empirical potential of the Finnis-Sinclair type proposed by Ackland et al \cite{Ackland:1995} for Zr-Zr potential. This Zr-Zr potential has been widely used to study the stable configurations and diffusion of SIAs in Zr \cite{Osetsky:2002, Woo:2003, Wen:2012, Diego:2011}. Although the most stable SIA configuration predicted by the potential is the basal-crowdion (BC), which deviates from the basal-octahedral (BO) predicted by the most recent ab initio modelling, the prediction that the basal-split (BS) configuration has the second-lowest energy agrees with the prediction of the ab initio modelling \cite{Peng:2012, Samolyuk:2014, Verite:2013, Varvenne:2013}. Comprehensively, the potential used in the present paper  makes predictions closer to those  of the newest ab initio modelling than those given by the potential developed later \cite{Mendelev:2007}.

The initial configurations of the simulation boxes were prepared by introducing a Zr atom to a Zr substrate in an hcp structure at a randomly selected position. For statistical purposes and to optimally use GPU-accelerated MD simulations \cite{Hou:2013}, we generated 1000 independent replicas of simulation boxes, each of which contained 6481 Zr atoms. The x, y and z axes were set along $[\overline{1}2\overline{1}0]$, $[\overline{1}010]$ and $[0001]$, and the periodic condition was applied in the three directions. The boxes were then thermalized and relaxed at a given temperature ranging from $300 K$ to $1200 K$ until thermal equilibrium was reached. The boxes were then relaxed further for $30 ps$. The trajectories of atoms were usually recorded every $0.1ps$. A finer recording time step was used only for observing the detailed transition path between SIA configurations. Because 1000 replicas of simulation boxes were used in parallel, the simulation procedure is equivalent to running one simulation box for $30 ns$.

In addition to visually observing the atomic configurations, quantitative analysis was conducted. Methodologically, correct identification of SIAs is essential for calculating the SIA diffusion properties, especially in cases of high temperature. As will be shown in our results, the BS configuration is the SIA state appearing most frequently in SIA migration. It is not meaningful to distinguish which atom in the atom pair of the BS configuration is the self-interstitial atom. Thus, in place of identifying the exact position a self-interstitial atom, we identify the location of a self-interstitial atom (LSIA). An LSIA is defined by the position of the hcp lattice point at which more than one atom fall in the Wigner-Seitz cell of the lattice point. A similar method was used by Osetsky et al \cite{Osetsky:2002} in their calculations. This method was also used to identify interstitials and vacancies produced in cascade collisions and was shown to be robust \cite{Nordlund:1998}. We also tested the robustness of the method for the MD simulations in the present paper. In a simulation box created above, only one LSIA can be found, and an LSIA contains only two atoms for all temperatures considered in the present paper. Another advantage of the method is that we can easily determine the moving direction of the LSIA, immune to the influence of temperature. Because of this, we can differentiate the types of events in SIA migration.

%%%%%%%%%%%%%%%%%%%%%%%%%%%%%%%%%%%%%%%%%%%%%%%%%%%%%%%%%%%%%%%%%%%
%-------------------------------------------------------------------------------
\subsection{Features of LSIA migration}
\label{subsec:MD_LSIA}

%-------------------------------------------------------
\begin{figure*}[thp!]
\includegraphics[width=12cm,clip]{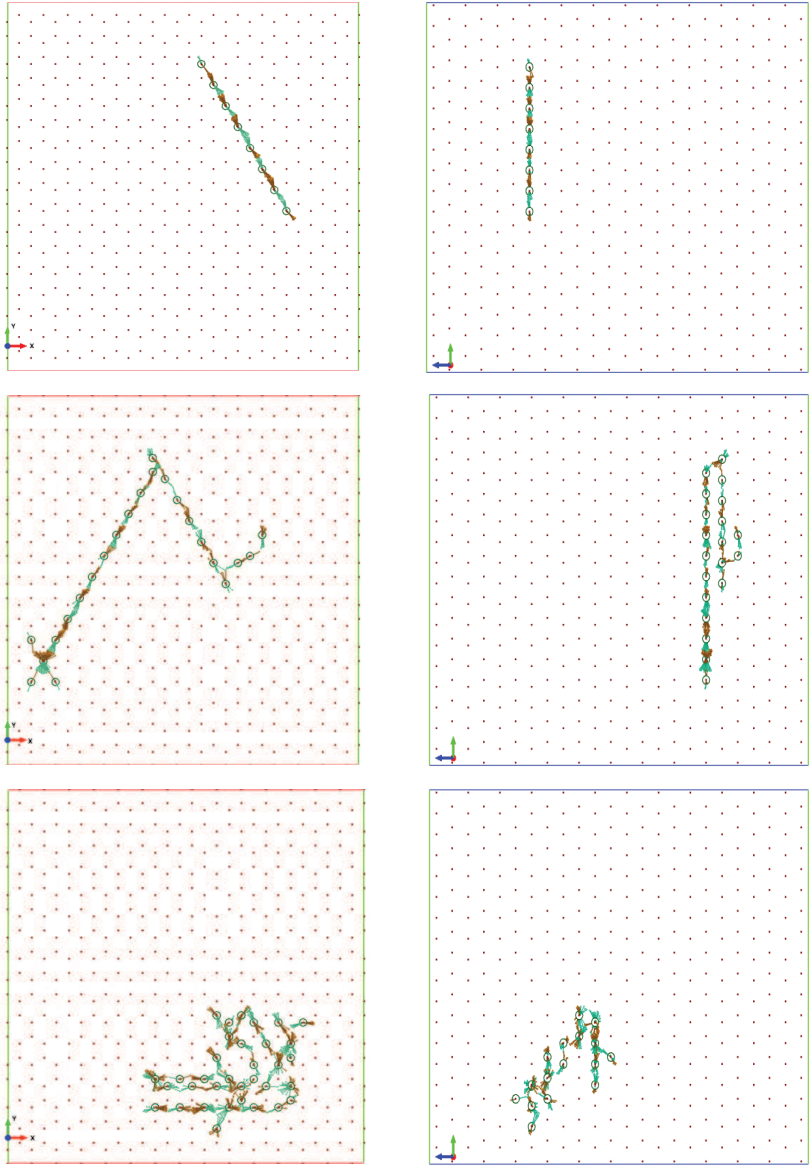}
\caption{(color online).
Typical trajectories of LSIAs for temperature $T=300K$, $600K$ and $900K$ (from top row to bottom row). Left column are views in the $[0001]$ direction, right column are views in the $[\overline{1}2\overline{1}0]$ direction. Dark red dot: hcp lattice point; Circle: LSIA; Gold dot and line: an interstitial atom and its connection with its LSIA; Green dot and line: the same as the gold dot and line but for another interstitial atom at the LSIA.
\label{fig:SIA_VISUAL}
}
\end{figure*}
%-------------------------------------------------------

\begin{figure}[h]
\includegraphics[width=8cm,clip]{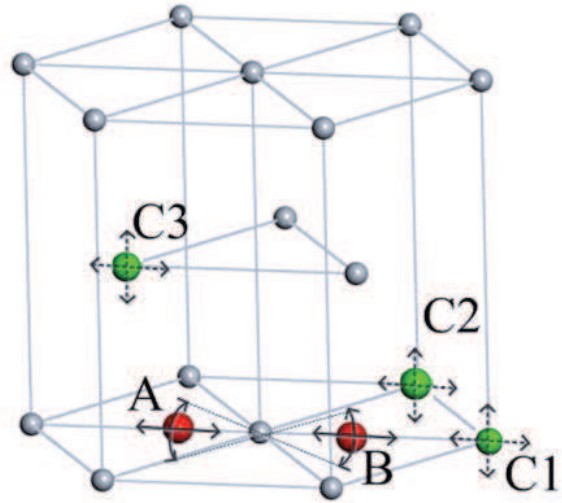}
\caption{(color online).
Schematic graph for the movement of interstitial atoms and LSIAs. A and B denote the pair of interstitial atoms, which swing and vibrate around their LSIAs. C1, C2 and C3 denote the atoms on three of the nearest lattices of the LSIA. These atoms on the LSIA candidates also swing and vibrate around their lattice points. When the interstitial atoms and the atom on one of the LSIA candidates approximately line up, a jump of the LSIA may occur, with one of the interstitial atoms falling into the new LSIA and another returning to lattice point where the previous LSIA was located.
\label{fig:JUMP_SCHAME}
}
\end{figure}

According to the simulation approach described above, MD simulations were performed at different substrate temperatures $T_{s}$. Fig. \ref{fig:SIA_VISUAL} displays the typical trajectories of LSIAs for $T_{s}=300 K$, $600 K$ and $900 K$. The graphs were created by merging the snapshots at different time steps. Also shown are the interstitial atoms and their connections with the LSIAs at which they are located. The first point to observe is that the pair of interstitial atoms in LSIAs are in BS or BC configuration for most of the time (it is actually difficult to distinguish BS and BC configuration for nonzero substrate temperatures). However, the atom-LSIA-atom connection is not necessarily a straight line in the $\langle11\overline{2}0\rangle$ direction. Fig. \ref{fig:JUMP_SCHAME} schematically displays the typical configurations of the interstitial atoms (denoted as A and B). The interstitial atoms exhibit superposition movements of thermal swinging and vibrating around their LSIA. The atoms (denoted as C1, C2 and C3) on the nearest lattice sites of the LSIA also thermally swing and vibrate around their lattice points. The nearest lattice sites of the LSIAs are LSIA candidates. There are moments when the interstitial atoms and the atom on one of the LSIA candidates approximately line up. At these moments, an LSIA jump may occur, during which one of the interstitial atoms falls in the new LSIA and another  returns to the lattice point where the previous LSIA was located. When we quenched the simulation boxes to zero temperature, the SIA configuration denoted as PS¡¯ in reference \cite{Varvenne:2013}, which leads to the migration of SIAs from one basal plane to another, was indeed found. However, the appearance of this state is infrequent. Most of the time, the interstitial atoms tend to swing around their LSIAs at small angles. This leads to our second point to observe. From Fig. \ref{fig:SIA_VISUAL}, it is seen that the LSIA exhibits one-dimensional migration at low temperatures ($T_{s} = 300K$). The LSIA ¡®jumps¡¯ forward and backward in a $\langle11\overline{2}0\rangle$ direction. With increasing temperature ($T_{s} = 600K$), changes in the moving direction of LSIAs are observed. An LSIA may change its moving direction from one $\langle11\overline{2}0\rangle$ direction to another equivalent $\langle11\overline{2}0\rangle$ direction on the same basal plane. It may also first jump in one of the two nearest $\{0001\}$ planes and then quickly turn into movement in one $\langle11\overline{2}0\rangle$ direction. However, the probability of an LSIA changing its moving direction in-plane or jumping off-plane is significantly less than the probability of it continuing to move one-dimensionally. The trajectories of LSIAs are thus observed to be constituted by connected line segments of various lengths. With a further increase in temperature ($T_{s} = 900 K$), the probability of an LSIA changing its moving direction increases, and the average length of the line segments thus decreases. Even so, the events in which an LSIA jumps forward or backward in a $\langle11\overline{2}0\rangle$ direction still occur most frequently.

%-------------------------------------------------------
%-------------------------------------------------------

\begin{figure}[h]
\includegraphics[width=8.cm,clip]{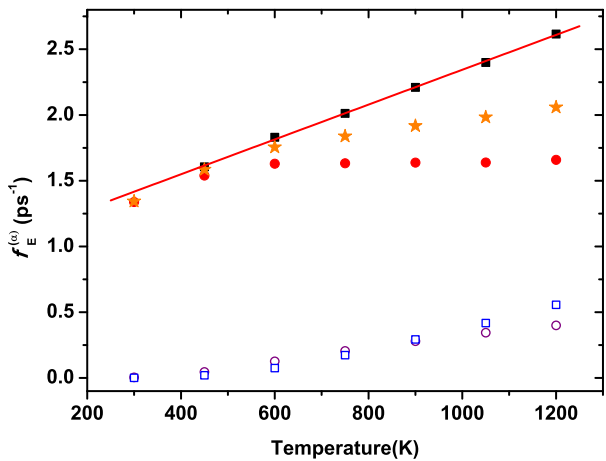}
\caption{(color online).
The occurrence frequencies of event types, $f_{E}^{(\alpha)}$, as a function of temperature, where solid square denotes the total frequency of events $f_{E}^{(S)}$; solid circle denotes $f_{E}^{(ILJ)}$; hollow circle denotes $f_{E}^{(OLJ)}$; hollow square denotes $f_{E}^{(OPJ)}$; and solid star denotes $f_{E}^{(IPJ)}=f_{E}^{(ILJ)}+f_{E}^{(OLJ)}$. The red solid line is the linear fitting of $f_{E}^{(S)}$: $f_{E}(T_{s})=1.01906 T_{s}+0.00133$, with $T_{s}$ in $K$ and $f_{E}(T_{s})$ in unit $ps^{-1}$.
\label{fig:JUMP_FRE}
}
\end{figure}

To conduct a more quantitative analysis of the migration feature of Zr SIA, we calculated occurrence frequencies of events, $f_{E}^{(\alpha)}$, where the subscript ¦Á denotes the event type. The occurrence of an event is determined if an LSIA changes. If $N_{E}^{(\alpha)}$ is the number of times event $\alpha$ occurs in all 1000 simulation boxes in the simulation time of $30 ps$, then $f_{E}^{(\alpha)}$ in unit $s^{-1}$ is calculated by $(10^{8}/3)N_{E}^{(\alpha)}$. We differentiated three types of events: in-line jump (ILJ), off-line jump (OLJ), and off-plane jump (OPJ), as schematically displayed in Fig .\ref{fig:JUMP_SCHAME}. An off-plane jump is easily identified when an LSIA changes from one basal plane to another. A jump is identified as off-line if the LSIA jumps in the same basal plane but in a different direction from its intermediately previous jump in a $\langle11\overline{2}0\rangle$ direction. Otherwise, a jump is an in-line jump. It should be mentioned that in a recording time step ($0.1ps$)  only the LSIA jumps between two adjacent hcp lattice sites were observed. Fig. \ref{fig:JUMP_FRE} shows $f_{E}^{(\alpha)}$ as a function of the temperature. Also shown is the summation of the occurrence frequencies of all types of events, $f_{E}^{(S)}$. It is seen that the in-line jump frequency $f_{E}^{(ILJ)}$ is significantly higher than the off-line jump frequency $f_{E}^{(OLJ)}$ and off-plane jump frequency $f_{E}^{(OPJ)}$ in the considered range of the temperatures. At room temperature ($T_{s} = 300 K$), $f_{E}^{(OLJ)}$ and $f_{E}^{(OPJ)}$ are very small compared with $f_{E}^{(ILJ)}$. With increasing temperature in the range of $300 K \leq T_{s} \leq 600 K$, $f_{E}^{(ILJ)}$ increases but at a decreasing rate. For $T_{s}\geq 600 K$, $f_{E}^{(ILJ)}$ tends to be weakly dependent on the temperature. Conversely, $f_{E}^{(OLJ)}$ and $f_{E}^{(OPJ)}$ increase with increasing temperature at an increasing rate in the range of $300 K \leq T_{s} \leq 600 K$. For $T_{s} \geq 600 K$, $f_{E}^{(OLJ)}$ and $f_{E}^{(OPJ)}$ tend to linearly increase with increasing temperature. $f_{E}^{(OPJ)}$ increases slightly faster than $f_{E}^{(OLJ)}$ and slightly overtakes $f_{E}^{(OLJ)}$ for $T_{s} \geq 900 K$. The increasing rate of $f_{E}^{(OLJ)}$ and $f_{E}^{(OPJ)}$ increase compensates the decreasing increase of $f_{E}^{(ILJ)}$. This leads to good linear dependence of the total occurrence frequencies of all events, $f_{E}^{(S)}$, on the temperature. Using the data of $f_{E}^{(S)}$ for $T_{s} \geq 600 K$, the dependence of $f_{E}^{(S)}$ on the temperature was very well fitted by equation $f_{E}(T_{s}) = \mu T_{s}+\mu_{0}$ with $\mu=1.01906 K^{-1} ps^{-1}$ and $\mu_{0}=0.00133 ps^{-1}$. The equation $f_{E}(T_{s})$ can be applied even for lower temperature, with a small deviation at $T_{s} = 300 K$. In contrast, dependence of $f_{E}^{(S)}$ on the temperature cannot be well fitted by a single Arrhenius relation in the whole temperature range considered. This result suggests that the jump frequency of all events follows the Einstein-Smoluchowski relationship better than the Arrhenius relationship and that the SIA migration is thus more Brownian-like when the temperature is higher than room temperature, although the occurrence frequency for a specific type of event, for example $f_{E}^{(ILJ)}$, may not always follow the Einstein-Smoluchowski relationship in the considered temperature range.
%-------------------------------------------------------
%-------------------------------------------------------
For the purpose of comparing our results  with those of other authors, the the occurrence frequency of in-plane jump, $f_{E}^{(IPJ)}=f_{E}^{(ILJ)}+f_{E}^{(OLJ)}$, is also shown in Fig. \ref{fig:JUMP_FRE}. In contrast to $f_{E}^{(S)}$, the dependence of the in-plane jump frequency $f_{E}^{(IPJ)}$ on $T_{s}$ should be considered in two temperature regimes divided by $T_{s} = 600 K$. This is in agreement with what was observed in the work of Osetsky et al \cite{Osetsky:2002}, but for unknown reason, the in-plane jump frequency displayed in reference \cite{Osetsky:2002} is systematically larger than ours by a factor of approximately two. Osetsky et al had fitted in-plane jump frequency to the Arrhenius relationship in two temperature regimes ($T_{s} \leq 600 K$ and $T_{s} \geq 800 K$), with the activated energy equal to $0.007 eV$ and $0.028 eV$ , respectively. The small activated energy in the low temperature regime ($0.007 eV$), which is comparable to or even smaller than the considered temperature, is actually an indication that the SIA migration tends to be Brownian-like rather than activated migration.

\begin{figure}[h]
\includegraphics[width=8.cm,clip]{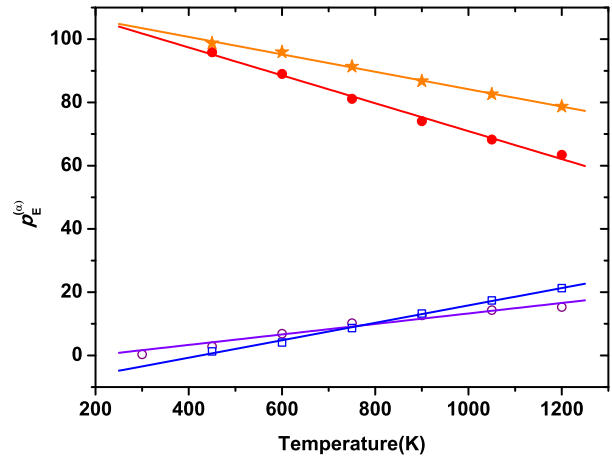}
\caption{(color online).
The percentage of occurrence frequencies of event types, $p_{E}^{(\alpha)}=f_{E}^{(\alpha)}/f_{E}^{(S)}$, as a function of temperature. solid circle denotes $p_{E}^{(ILJ)}$; hollow circle denotes $p_{E}^{(OLJ)}$; hollow square denotes $p_{E}^{(OPJ)}$; solid star denotes $p_{E}^{(IPJ)}$. The solid lines are the linear fittings of $p_{E}^{(\alpha)}$: $p^{(\alpha)}(T_{s})=A^{(\alpha)} T_{s}+B^{(\alpha)}$, with the fitting parameters given in Table \ref{tab:Fitting_para}.
\label{fig:JUMP_PECENTAGE}
}
\end{figure}

Fig. \ref{fig:JUMP_PECENTAGE} displays the percentage of occurrence frequencies of event types, defined by $p_{E}^{(\alpha)}=f_{E}^{(\alpha)}/f_{E}^{(S)}$. Again, $p_{E}^{(\alpha)}$ can be fitted by the linear equation $p^{(\alpha)}(T_{s})=A^{(\alpha)} T_{s} + B^{(\alpha)}$, especially for $T_{s} \geq 600 K$. The fitting parameters $A^{(\alpha)}$ and $B^{(\alpha)}$ obtained by fitting to $p_{E}^{(\alpha)}$ data of $T_{s} \geq 600 K$ are given in Table \ref{tab:Fitting_para}. The percentage of in-line jump frequency $p_{E}^{(ILJ)}$ is in good agreement with $p^{(ILJ)}$ event for the low temperature. Some deviations from the linear equation can be observed for $p_{E}^{(OLJ)}$ and  $p_{E}^{(OPJ)}$ in the low temperature regime. However, because $p_{E}^{(OLJ)}$ and  $p_{E}^{(OPJ)}$ are small at low temperatures, the deviations are less important from the viewpoint of applications such as KMC in which the occurrence frequency of type $\alpha$ is calculated by $f_{E}(T_{s})\cdot p^{(\alpha)}(T_{s})$. Again, it is clearly seen that the in-line jumps are the major events in LSIA migration even at $T_{s} = 1200 K$ or higher.

%-------------------------------------------------------
%-------------------------------------------------------

%------------------------------------------------------------------------
\begin{table}[h]
\caption{\label{tab:Fitting_para}
The fitting parameters of fitting $p^{(\alpha)}(T_{s})=A^{(\alpha)} T_{s}+B^{(\alpha)}$ to the percentage of occurrence frequencies of event types in Fig.\ref{fig:JUMP_PECENTAGE}
}
\begin{ruledtabular}
\begin{tabular}{rrrrrrr}
Event type    &~~ILJ	&~OLJ &~IPJ &~OPJ \\
\hline
%\multicolumn {5}{c}        \\
$A^{(\alpha)}(K^{-1}) $ & $~~-0.0441$	&$~0.0166$	&$~-0.0275$	&$~0.0275$\\
$B^{(\alpha)}$         &  $~~115.0209$ &$~-3.3084$ &$~111.7124$ &$~-11.7124$\\
\end{tabular}
\end{ruledtabular}
\end{table}

For the convenience of description, we call the migration, which is featured by migration trajectories that are constructed by line segments of temperature-dependent average lengths in hcp-basal planes, fraction-dimensional migration. The kinetic effects of the fraction-dimensional migration could be different from those that would occur if there was only conventional one-dimensional or two-dimensional migration in the basal plane, which will be defined in the next section. In the next section, based on Monte Carlo simulations, we account for the kinetic effects in which the feature of the fraction-dimensional migration may results.

%%%%%%%%%%%%%%%%%%%%%%%%%%%%%%%%%%%%%%%%%%%%%%%%%%%%%%%%%%%%%%%%%%%
%==================================================================
\section{Visited sites by SIAs}
\label{sec:MC simulation}
%------------------------------------------------------------------

As  demonstrated in the previous section, the migration of LSIA in the basal plane of Zr is fraction-dimensional. According to the random walk theory \cite{Chandrasekhar:1943}, a particle randomly walking on one-dimensional lattice sites can repeatedly visit the lattice sites many times. If off-line jumps occur, the visits on these sites are branched, and sites aligned in other directions, which could not be visited when the particle walks in one direction, can probably be visited. We suggest to use the average number of visited sites per event, $n_{SPE}$ , to measure the potential kinetic effects. From the viewpoint of applications, the interaction rate between SIAs or SIAs and defects of other types (e.g., vacancies) should be proportional to $n_{SPE}$.

We performed Monte Carlo (MC) simulations to calculate $n_{SPE}$. We started an MC simulation from a randomly selected hcp lattice site. The next site to be visited was randomly chosen from the candidate sites of an event, with the type of event determined by sampling according to the percentage of occurrence frequencies $p_{E}^{(\alpha)}$. The numbers of candidate sites were 2, 4 and 6 for ILJ, OLJ and OPJ, respectively. A site can contribute only one count to the number $N_{V}$  of visited sites. The simulation terminates after issuing an assumed number $N_{E}$  of events. $N_{E}$ can be considered as the average number of jumps for an LSIA moving before its annihilation due to, for example, recombination between SIAs and vacancies. $N_{E}$  can also be converted to migration time $t$ by $t=N_{E}/f_{E}^{(S)}$, where $f_{E}^{(S)}$ is the occurrence frequency of all event types. For a given $N_{E}$ , many MC simulation runs were performed. $n_{SPE}$  was then calculated by $\langle N_{V}\rangle / N_{E}$ , with $\langle \rangle$ denoting the average over the runs. The diffusion coefficient $D^{(2)}$ in the basal-plane was calculated by $\langle x^{2}+y^{2}\rangle /4t$, and the three-dimensional diffusion coefficient $D^{(3)}$ was calculated by $\langle x^{2}+y^{2}+z^{2} \rangle/6t$, with $(x,y,z)$ denoting the displacement vectors of LSIAs after $N_{E}$ ($\gg 1$) jumps.

\begin{figure}[h]
\includegraphics[width=8.5cm,clip]{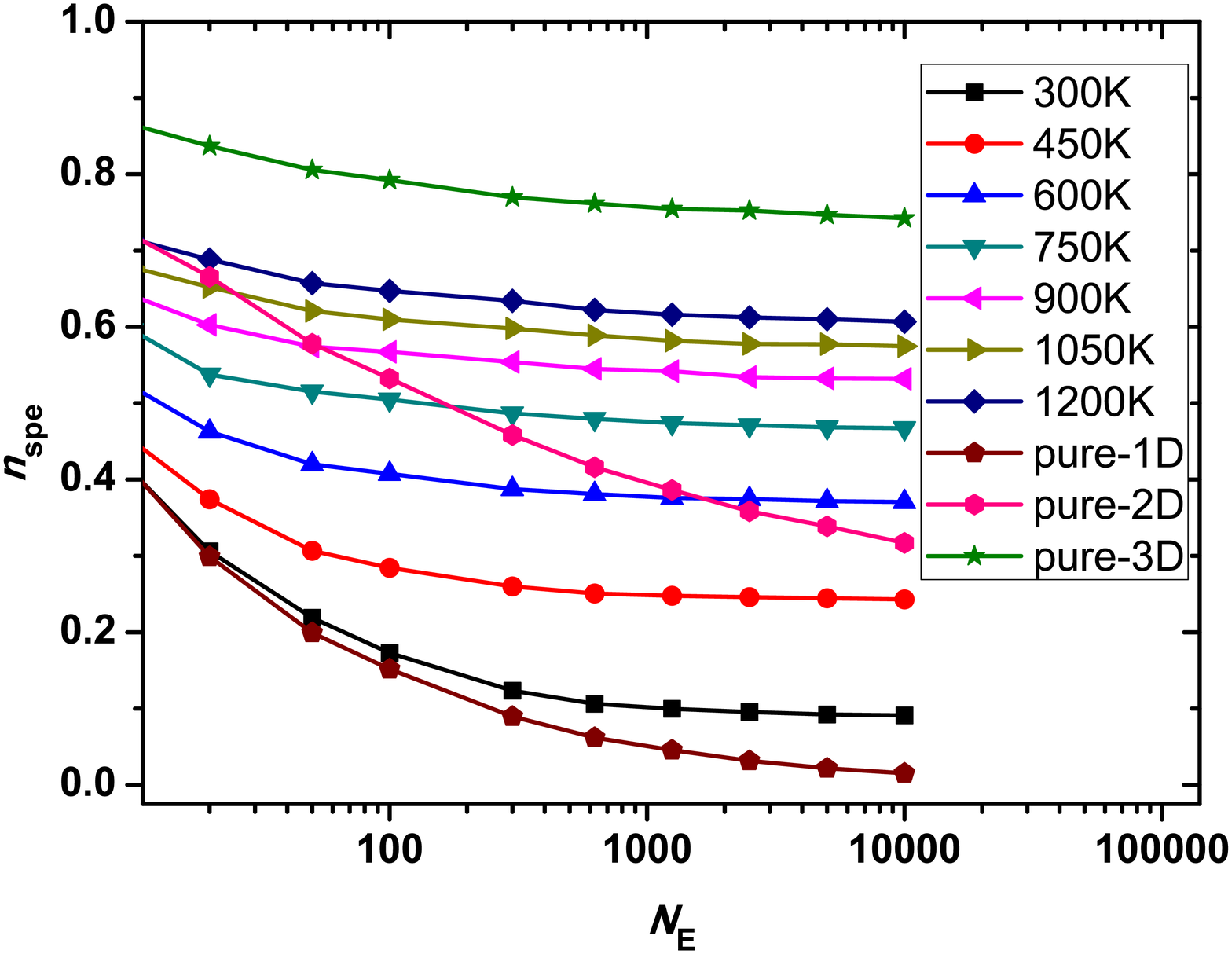}
\caption{(color online).
The average number of visited sites per event, $n_{SPE}$ , as a function of the number of jump events $N_{E}$. Also shown are $n_{SPE}$ for full three-, two- and one-dimensional migration. The definitions of full three-, two- and one-dimensional migration can be found in the text.
\label{fig:NSPE_NE}
}
\end{figure}

Fig. \ref{fig:NSPE_NE} displays $n_{SPE}$  as a function of $N_{E}$ , obtained using $p_{E}^{(\alpha)} (T_{s})$ given in subsection~\ref{subsec:MD_LSIA} for different substrate temperatures. Also shown in Fig. \ref{fig:NSPE_NE} is $n_{SPE}^{(d)}$ for full three- and two- and one-dimensional migration of LSIA on hcp-lattices, where the superscript $d$ denotes the dimensionality. Here, we denote ¡®full¡¯ three-dimensional migration as the case in which the probabilities of jumps from a site to all of its 12 nearest neighboring sites are the same, ¡®full¡¯ two-dimensional migration as the case in which the probabilities of jumps from a site to its 6 nearest neighboring sites in the same plane are the same, and ¡®full¡¯ one-dimensional migrations as the case in which only in-line jumping exists. Obviously, two- and one-dimensional migrations are special cases of the fraction-dimensional migration. Keeping in mind that mathematically, $n_{SPE}=1$ if $N_{E} =1$ in any case, all $n_{SPE}$ values are seen to be monotonically decreasing with increasing $N_{E}$  due to the spreading out of the visited lattice sites. The migration dimensionality is lower, the level of the spreading out is less. Thus, $n_{SPE}^{(1)}$ and $n_{SPE}^{(3)}$ provide the minimum and maximum boundaries for $n_{SPE}$  of the migration phenomenon in the hcp-lattice, respectively. The difference between the minimum and maximum is $53.8\%$ at $N_{E} =20$. The difference increases with increasing $N_{E}$ . The large difference suggests that $n_{SPE}$  is very sensitive to the migration dimensionality. Returning to the migration of LSIA in Zr, it is seen that $n_{SPE}$  varies from $30.6\%$ to $68.8\%$ at $N_{E} =20$ and from $9.1\%$ to $60.7\%$ at $N_{E} =10,000$ , when the temperature $T_{s}$  increases from $300 K$ to $1200 K$. Another observation is that the increase of $p_{E}^{(OPJ)}$ causes $n_{SPE}$  to approach a constant asymptotically for large $N_{E}$. For full two-dimensional and one-dimensional migration, the slope of $n_{SPE}$  as a function of $N_{E}$ is large even for $N_{E} =1000$. The large slope is probably an indication that the kinetics of the LSIAs could be dependent of the concentration of the sites with which the SIAs could interact. In such a case, the use of concentration-independent kinetic quantities, such as the diffusion coefficient that originates from the theory of continuous diffusion, should be revalidated. We will address this issue more generally in the future. Here, we address another question on the usage of diffusion coefficients when the LSIA migration is fraction-dimensional.
%-------------------------------------------------------
%-------------------------------------------------------

\begin{figure}[h]
\includegraphics[width=8.5cm,clip]{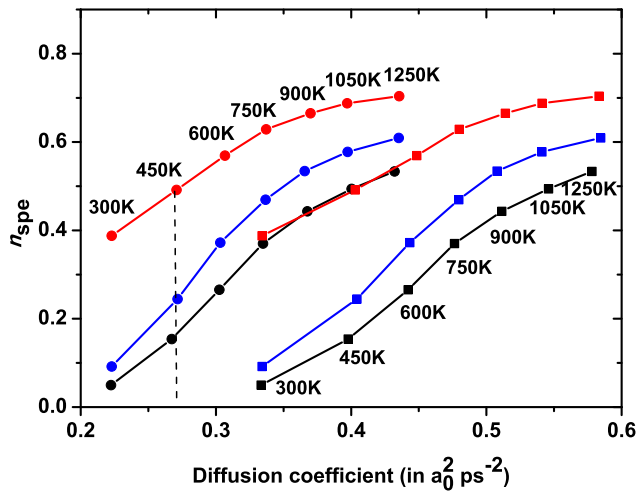}
\caption{(color online).
$n_{SPE}$ vs diffusion coefficients $D^{(2)}$(denoted by solid square) and $D^{(3)}$(denoted by solid circle) for: two-dimensional case (red); fraction-dimensional case (blue); and one-dimensional case (black) (see text for definitions of the cases).
\label{fig:NSPE_D}
}
\end{figure}

Because an LSIA jump in the basal plane always contributes the same square displacement $a_{0}^{2}$, where $a_{0}$  is the basal lattice length, to the mean square displacement (MSD) whether the jump is in-line or off-line, one cannot actually judge solely by the diffusion coefficient whether  the migration of SIAs in the basal plane is one-dimensional, two-dimensional or fraction-dimensional. The in-basal-plane migration of an LSIA in Zr is usually treated either as one-dimensional or two-dimensional \cite{Woo:2000, Arevalo:2007, Christien:2005, Samolyuk:2014}. However, our present results reveal that the in-basal-plane migration of LSIA in Zr is fraction-dimensional. To study the impacts of such treatments, we performed further MC simulations for three cases. The first case is the ¡®real¡¯ fraction-dimensional case, in which all $p_{E}^{(¦Á)} (T_{s} )$ are what are obtained by MD simulations in subsection~\ref{subsec:MD_LSIA}. For the other two cases, the same $p_{E}^{(OPJ)}(T_{s})$ was also adopted. However, we set $p_{E}^{(ILJ)} (T_{s} )=1-p_{E}^{(OPJ)} (T_{s})$ and $p_{E}^{(OLJ)} (T_{s})=0$ to treat the in-basal-plane migration of LSIA as full one-dimensional, and set $p_{E}^{(ILJ)} (T_{s})=2(1-p_{E}^{(OPJ)} (T_{s}))/6$ and $p_{E}^{(OLJ)}(T_{s})=4(1-p_{E}^{(OPJ)} (T_{s}))/6$, to treat the in-basal-plane migration of LSIA as full two-dimensional. Fig. \ref{fig:NSPE_D} shows $n_{SPE}$ versus diffusion coefficient $D^{(2)}(T_{s})$ and $D^{(3)}(T_{s})$, obtained with $N_{E} =10,000$ for all three cases. The temperature $T_{s}$ is from $300 K$ to $1200 K$. As expected, at the same temperature, the three cases have the same diffusion coefficients $D^{(2)}(T_{s})$ and $D^{(3)}(T_{s})$ in statistical accuracy, as denoted by the dashed line in Fig. \ref{fig:NSPE_D}. However, the $n_{SPE}$  of the ¡®real¡¯ case deviates significantly from the $n_{SPE}$  of the assumed one- and two-dimensional cases. This result suggests that the conventional diffusion coefficients cannot reflect the real underlying kinetics of LSIA migration in Zr, in which the in-basal-plane migration of LSIAs is neither fully one-dimensional nor fully two-dimensional but rather fraction-dimensional. Thus, the models in downstream applications, such as KMC or mead field rate theory, for modeling the microstructure evolution in Zr should account for the effects of the fraction-dimensional migration of SIAs to make accurate predictions.

%-------------------------------------------------------
%-------------------------------------------------------

%%%%%%%%%%%%%%%%%%%%%%%%%%%%%%%%%%%%%%%%%%%%%%%%%%%%%%%%%%%%%%%%%%%
%==================================================================
\section{Conclusions}
\label{sec:Conclusions}
%------------------------------------------------------------------
Based on MD simulations, we have analyzed the migration of LSIAs in Zr by extracting the occurrence frequencies of ILJ, OLJ and OPJ. On the whole, the migration of LSIAs tends to be Brownian-like, although the three types of jumps may not follow Brownian and Arrhenius behavior. Because the occurrence frequencies of ILJ, OLJ and OPJ are equal, with ILJs being dominant, the trajectories of LSAs in the hcp basal-plane exhibit the feature of fraction-dimension. Based on MC simulations, we have analyzed the potential kinetic impacts of fraction-dimensional migration, which is measured by $n_{SPE}$, the average number of lattice sites visited by a jump event. The $n_{SPE}$ of fraction-dimension migrations could be very different from that if the migration in hcp basal-plane is conventional one- or two-dimensional. This result suggests that the conventional diffusion coefficient calculated by mean square displacements of particles cannot reflect the migration dimensionality and thus cannot provide an accurate description of the underlying kinetics of SIAs in Zr. This conclusion may not be limited to SIA migration in Zr and could be more generally meaningful to the situations  in which low dimensional migration of defects in materials have been observed.

%%%%%%%%%%%%%%%%%%%%%%%%%%%%%%%%%%%%%%%%%%%%%%%%%%%%%%%%%%%%%%%%%%%%%%%%%%%%%%%%%%%%%%
%=====================================================================================
\section{Acknowledgments}
%--------------------------------------------------------------------------------------
This work was partly supported by the National Natural Science Foundation of China (Contract Nos. 91126001) and the National Magnetic Confinement Fusion Program of China (2013GB109002). The authors thank Dr. Ziwen FU for his help during the preparation of the manuscript.

%%%%%%%%%%%%%%%%%%%%%%%%%%%%%%%%%%%%%%%%%%%%%%%%%%%%%%%%%%%%%%%%%%%%%%%%%%%%%%%%%
%////////////////////////////////////////////////////////////////////////////////
\bibliography{mybibfilea}
%==================================================================
\end{document}